\documentclass[onecolumn,linenumbers, showpacs]{revtex4}
\usepackage{graphicx}
\usepackage{graphicx}
\usepackage{dcolumn}
\usepackage{bm}
\usepackage[latin1]{inputenc}
\usepackage{amsmath}
\usepackage{amsfonts}
\usepackage{amssymb}
\usepackage{mathrsfs}
\RequirePackage[colorlinks,citecolor=blue,urlcolor=magenta,linkcolor=blue]{hyperref}

\input epsf

\begin{document}

\title{\Large Complete cosmic scenario in the Randall-Sundrum braneworld from the dynamical systems perspective}
\smallskip

\author{\bf~Jibitesh~Dutta $^{1,2}$\footnote{jdutta29@gmail.com,~jibitesh@nehu.ac.in}
 and H. Zonunmawia $^2$\footnote{zonunmawiah@gmail.com}, }

\smallskip

\affiliation{$^{1}$Mathematics Division, Department of Basic
Sciences and Social Sciences,~ North Eastern Hill University,~NEHU
Campus, Shillong - 793022 ( INDIA )}

\affiliation{$^2$Department of Mathematics,~ North Eastern Hill
University,~NEHU Campus, Shillong - 793022 ( INDIA )}

\date{\today}

\begin{abstract}
 The paper deals with dynamical system analysis of a coupled scalar field in the Randall-Sundrum(RS)2 brane
world. The late time attractor describes the final state of the
cosmic evolution. In RS2 based phantom model there is no late-time
attractor and consequently there is uncertainty in cosmic
evolution. In this paper, we have shown that it is possible to get
late-time attractor when gravity is coupled to scalar field.
Finally, in order to predict the final evolution of the universe,
 we have also studied classical stability of the model. It is found that there are late time attractors
 which are both locally  as well as classically stable and so our model can realise the
 late time cosmic acceleration.

\end{abstract}


\maketitle



\section{Introduction}

  The modified theories of gravity which can explain the present observed accelerated expansion of the
  universe \cite{sp,sp03,ag} without any need of dark energy(DE) \cite{ejc,kbamba} have received lot of success  in
  recent years. Brane gravity(BG)/Braneworld scenario inspired by advances of String/M-Theory is a prominent theory of  modified
  gravity. In this scenario our observable universe is assumed to be a sub-manifold embedded in compactified  higher
  dimensional space time  known as bulk. Here except gravity, all the standard model particles are confined to sub-
  manifold called brane. Braneworld models provides a novel way to correct General Relativity(GR) and gives answer
  to several unsolved problems of cosmology \cite{rm}.

   Among different braneworld models, the Randall-Sundrum type II model(RS2)
  is very popular among cosmologists due to its  rich conceptual base \cite{rs2}. RS2 model gave a
  framework often referred as braneworlds with wrapped extra-dimension in contrast to Kaluza-Klein mechanism of
  compactification where extra-dimension is very small \cite{rs1}. In this set up big-rip singularity of phantom
  model can be avoided and acceleration is transient in nature \cite{sksjd}. In contrast to GR
  based models here inflation is possible for a large class of potentials  due to modified Friedmann equation\cite{rmhawkins}.

 Another braneworld model that received considerable attention is the
Dvali-Gabadadze-Porrati(DGP) braneworld model
\cite{dgp1,dgp2,dgp3}. This model can explain the present
acceleration without using mysterious DE. Usually RS2 model
modifies GR at small cosmological scales and DGP model modifies GR
at large cosmological distance. Consequently RS2 models have more
impact on cosmology of early universe. The interesting and recent
feature is that RS2 model can also modify
 late time cosmic expansion if  the energy density of the matter content increases at late time
 \textit{e.g.}, phantom scalar field \cite{sksjd, GarciaSalcedo:2010wa}. Thus for scalar fields like phantom,
 RS2 models can have impact on late universe and it can give a complete cosmic scenario.

Dynamical systems tools have a central role for qualitative
analysis of homogeneous cosmological models whose evolution is
governed by finite dimensional autonomous system of ordinary
differential equations (ASODE) \cite{dsincosmology,Coley:2003mj}.
The goal of the dynamical systems approach is to correlate
asymptotic cosmological solutions with important concepts like
past and future attractors along with saddle points. In the phase
space of models, if a critical point can be associated with a
solution, it means that in the neighbourhood of this solution the
universe  dynamics  will evolve for a sufficiently long time
irrespective of initial data. To describe a complete cosmological scenario, a cosmological model
described by an autonomous system must have a past time attractor
(to represent inflationary epoch), one or two saddle points (to
represent radiation  and matter dominated universe) and an late
time attractor (to represent accelerated expansion universe)
\cite{Boehmer:2014vea}.

 In cosmology scalar fields play an active role to model DE, inflation and other
 cosmological features \cite{ar,jm}. The dynamical study of a scalar field coupled with perfect
 fluid has found many applications
 in GR context. For a nice introduction  \cite{Boehmer:2014vea,ir1} and application of
  this technique in GR based cosmology one can see recent papers  \cite{nroy1,nroy,nrn1}. In RS2 model, using dynamical system approach
  Rudra $et$ $al$ investigated role of modified Chaplyging gas in explaining accelerated
  expansion of the universe
   \cite{prudra}. They showed that around fixed points universe expands with  a power-law. A dynamical study of
   Dirac Born Infield (DBI) trapped in the braneworld is explored
   in \cite{arx} and it is found that in ultra relativistic
   regime scalar field dominated and matter scaling
   solutions  exist. In braneworld context
   dynamics of self-interacting
scalar field trapped in a RS2 model for a wide variety of
potentials  have been studied in
\cite{Gonzalez:2008wa,Leyva:2009zz,de}.  Escobar $et$ $al$ did a
refined study of a scalar field trapped in RS2 model using Centre
Manifold Theory with an arbitrary potential. They proved that
there are no late time attractors with 5D modifications \cite{de}.
Furthermore, in a recent paper Biswas $et$ $al$ studied
interacting dark energy in RS2 braneworld using dynamical systems
tools \cite{Biswas:2015zka}. All the above studies were done for a
quintessence scalar field. It is found that only early times
cosmology is modified with respect to GR based models. On the
contrary for usual phantom scalar field, the higher dimensional
Brane effects of RS2 usual phantom scalar field model modifies
past and future asymptotic properties so that we get past time
attractor and saddle point only \cite{ GarciaSalcedo:2010wa}. In
GR based phantom scalar field model we get saddle point and late
time attractor but it does not admit past time attractor
\cite{GarciaSalcedo:2010wa,BG}.

In this paper we shall show that in RS2 model when gravity is
coupled to scalar field with scalar coupling function and a
potential, the late universe is also modified and we get a
complete cosmic scenario. The coupling function additionally helps
to control the behaviour of scalar field and have been used
extensively in unifying phantom inflation with late time
acceleration \cite{kbamba,sn,sc,as}. Recently Mahata $et$ $al$
\cite{Mahata:2013oza}  studied the dynamics of phantom
cosmological model where gravity is coupled to
 scalar field having scalar coupling function
in GR context using dynamical system tools. The aim of this paper
is to study  the dynamics of coupled scalar field   on RS2
  braneworld by using dynamical system
  tools.  Furthermore,
  we have also examined the classical stability of the model in addition to calculating
  various observable quantities in the asymptotic solutions. In contrast to Ref  \cite{Mahata:2013oza},
  we shall show that coupled scalar field give rich dynamics and late time acceleration solutions  are also classically stable.

The paper is organised as follows: The basic  and  the evolution
equations of the  autonomous system  are presented  in section (\ref{II}).
Sections (\ref{III}) and (\ref{IV}) deals with the analysis of the critical points
of the system in 4D and 3D respectively. The  classical stability of  the model has been
studied in section (\ref{V}). Final section is devoted to discussion and conclusion.

\section {Basic equations}\label{II}
\noindent The total action of the field equations in RS2 model is
written as
\begin{equation}\label{1}
S=S_{RS}+S_{\phi}+S_m
\end{equation}
 \noindent where
 \begin{equation}\label{2}
 S_{RS}=\int d^{5}x~\sqrt{-g^{(5)}}~(2R^{(5)}+\Lambda_5)+\int d^{4}x~\sqrt{-g}~\lambda
 \end{equation}
\noindent  is the action of RS2 model, ~$g^{(5)}_{\mu\nu}$ and
$g_{\mu\nu}$ are the bulk metric and the brane metric
respectively, $R^{(5)}$ is the Ricci scalar in the bulk and
 $\Lambda_5$ is the bulk cosmological constant, $S_\phi$ is the action of the scalar field coupled to gravity given by
  \begin{equation}\label{3}
  S_\phi=\int d^{4}x\sqrt{-g}~ \Big[~-\frac{1}{2}\mu(\phi)g^{\mu\nu}(\nabla_{\mu}\phi)(\nabla_{\nu}\phi)-V(\phi)~\Big]
   \end{equation}
  $\mu(\phi)$ is the coupling parameter, $V(\phi)$ is the potential of the scalar field
  and  $S_m=\int d^{4}x~\sqrt{-g}~L_m$
is the action of the matter field which is chosen as the cold dark
matter (CDM).

Observations support homogeneous and isotropic model of the late
universe, given by the line-element \cite{ad}
\begin{equation}\label{4}
  ds^2 = dt^2-a^{2}(t)[dx^2 +dy^2 + dz^2]
\end{equation}
\noindent where $a(t)$ is the scale factor. \noindent In this
space-time, from action (\ref{1}) we get the following RS2 model
based modified Friedmann field equations  \cite{rm}
\begin{equation}\label{5}
3H^2=\rho_T\Big(1+\frac{\rho_T}{2\lambda}\Big)
\end{equation}

  Here $ \lambda$ is the brane tension and  $\rho_T =\rho_\phi + \rho_m$ , $\rho_\phi$  and
  $ \rho_{m} $ are being the energy density of CDM and the scalar field
  respectively. It should be noted that we have neglected the
  cosmological constant in the $4$D brane  by choosing RS fine
  tuning condition and for simplicity of calculation we have chosen AdS bulk where dark radiation term vanishes.

  When gravity is coupled to the scalar field  with scalar coupling function, the energy density and thermodynamic pressure are given by

  \begin{equation}\label{6}
   \rho_{\phi}= \frac{1}{2}\mu(\phi)\dot {\phi}^2 +V(\phi)
  \hspace{1.5cm} \textrm{and} \hspace{1.5cm} p_{\phi}=\frac{1}{2}\mu(\phi)\dot {\phi}^2 -V(\phi)
  \end{equation}

\noindent where $\mu(\phi)$ is the coupling parameter chosen for arbitrary function of $\phi$ and $V(\phi)$ is the potential of the scalar field.

  \par Here we assume that the scalar field does not interact with CDM  and the energy conservation
  relations takes the form.
  \begin{equation}\label{7}
  \dot{\rho_\phi}+3H(\rho_\phi+p_\phi)=0
  \end{equation}

   \begin{equation}\label{8}
   \dot{\rho_m}+3H\rho_m=0
   \end{equation}\\

  \noindent Using Eqs.(\ref{5})~,~(\ref{6})~,~(\ref{7})~and~(\ref{8}), we get
   \begin{equation}\label{9}
  2\dot H=-\Big(1+\frac{\rho_T}{\lambda}\Big)(\mu(\phi) \dot \phi^2+\rho_m)
  \end{equation}\\
  This implies
  \begin{equation}\label{10}
  \mu(\phi)\dot \phi^{2}=-\Big(\frac{2\dot H}{1+\frac{\rho_T}{\lambda}}+\rho_m \Big)
  \end{equation}\\
  As clear from Eq.(\ref{10}), when $\dot H$ is positive, we have phantom phase where $\mu(\phi)$ is negative and
  using conservation Eqs.(\ref{7}) and (\ref{8}), evolution equation of  the scalar field is given by

	 \begin{equation}\label{11}
   \mu(\phi){\ddot{\phi}}+\frac{1}{2}\frac{d\mu(\phi)}{d\phi}\dot{\phi^2}+3H\mu(\phi)\dot\phi+\frac{dV}{d\phi}=0
   \end{equation}

   \par Following \cite{Leyva:2009zz}, we introduce the following dimensionless phase space variables in order to build
	ASODE of the above cosmological equation.

   \begin{equation}\label{12}
   x\equiv \frac{\sqrt{\mu(\phi)}~\dot \phi}{\sqrt{6}~H}, \hspace{0.2cm}  y\equiv\frac{\sqrt{V(\phi)}}{\sqrt{3}~H} \hspace{0.2cm} , \hspace{0.2cm} z\equiv\frac{\rho_T}{3H^2} \hspace{0.2cm} \textrm{and} \hspace{0.2cm} s\equiv \frac{\sqrt{6}}{\phi}
   \end{equation}

After these choice of phase space variables, we can write
 following ASODE for above evolution equations

$$x'=-3x-\beta_1(s)~y^2 +\frac{3}{2}x\Big(\frac{2-z}{z}\Big)(z+x^2-y^2)$$
\\
$$y'=\beta_1(s)~xy +\frac{3}{2}y\Big(\frac{2-z}{z}\Big)(z+x^2-y^2)$$
\\
$$z'=3(1-z)+(z+x^2-y^2)$$
\\
\begin{equation}\label{13}
s'=-\frac{xs^2}{\sqrt{\mu_1(s)}}
\end{equation}
where $\beta_1(s)=\beta(\phi)=\sqrt{\frac{3}{2\mu(\phi)}}~\frac{d\ln V}{d\phi}$~,~$\mu_1(s)=\mu(\phi)$
and the primes denotes differentiation with respect to new time variable $U= \ln a$.
After the above choice of variables one can realise that
\begin{equation}\label{14}
\frac{\rho_T}{\lambda}=\frac{2(1-z)}{z},\;\Rightarrow\;0<z\leq 1
\end{equation}
 Eq.(\ref{14})  tells us that at $z=1$, brane effects vanish and GR is
recovered. It may be noted that at the point $z=0$ the ratio
$\frac{\rho_T}{\lambda}$ is undefined but as we  are interested in
asymptotic behaviour, we can study the limiting case $z\rightarrow
0$ separately.

 The effective equation of state for the scalar field (\textit{i.e.} $p_\phi=(\nu_\phi-1)\rho_\phi)$ is given by
\begin{equation}\label{15}
\nu_\phi=\frac{2x^2}{x^2+y^2}
\end{equation}

From the above dimensionless phase space variables the equation of state (EoS) parameter has the expression
 \begin{equation}\label{16}
 \omega_\phi=\frac{p_\phi}{\rho_\phi}=\frac{x^2-y^2}{x^2+y^2}
 \end{equation}

 The density parameter for the scalar field has the same expression as that of GR \cite{Mahata:2013oza}
 \begin{equation}\label{17}
\Omega_\phi=x^2+y^2
\end{equation}

 Also we have

  \begin{equation}\label{18}
\Omega_m=z-x^2-y^2 \hspace{1.2cm} \textrm{and} \hspace{1.2cm} q=-1-\frac{\dot H}{H^2}
\end{equation}
where $\frac{\dot H}{H^2}=-\frac{3}{2}\Big(\frac{2-z}{z}\Big)(z+x^2-y^2)$ so that the deceleration parameter $q$ has the expression

\begin{equation}\label{19}
q=-1+\frac{3}{2}\Big(\frac{2-z}{z}\Big)(z+x^2-y^2)
\end{equation}

 For numerical calculations we have estimated initial conditions in such a way that it is consistent
 with present observations (~$\Omega_m=0.27$ and the deceleration parameter $q=-0.53$ \cite{rg}~), so
 \begin{equation}\label{20}
 \Omega_\phi=x^2_0+y^2_0=0.73
 \end{equation}
 Using these values in the constraint Eq.(\ref{18}), we get $z_0=1$, where $x_0$,~$y_0$ and $z_0$ are the present
 values of $x$,~$y$ and $z$ respectively. Eq.(\ref{19}) and (\ref{20}) yields a lower bound set of present work
  $x_0=0.489$ and $y_0=0.7$.

  Here we see that the expressions of CDM energy density parameter and deceleration parameter involve
   brane effects and it differs from that of GR \cite{GarciaSalcedo:2010wa}.

The phase space for the autonomous dynamical system driven by the evolution Eq.(\ref{13}) can be defined as:
\begin{equation}\label{21}
\Psi=\{(x,y,z):0 \leq{x^2}+y^2\leq {z},-1\leq {x}\leq{1},0\leq {y},0<z\leq{1}\}\times \{s \in \mathbb{R}\}
\end{equation}

\section  { Critical points of 4D autonomous system }\label{III}

  In this section we will  analyse in detail the critical points
  of the autonomous system  corresponding to RS2 model and their
  stability properties. Let $y'=g(y)$ be a given non-linear autonomous system of differential equations with a
  critical point $y_0 \in \mathbb{R}^{n}$, where
  $y'=\frac{dy}{dt}$, $y\equiv (y_1,y_2,.......,y_n)\in \mathbb{R}^{n}$ and
   $g:\mathbb{R}^{n}\rightarrow\mathbb{R}^{n}$. The system can be
  linearised around the critical point as $u'=Au=Df(y_0)u$, where $u=y-y_0$, $A=Df(y_0)$ and
  $Df(y)=\frac{\partial g_i}{\partial y_j}$, $i,j=1,2,3,.......n$ and $A$ is the Jacobian
  matrix of the system. A critical point of the system is said to be hyperbolic if
  none of the eigenvalues of the Jacobian matrix $A$ have zero real part. If the Jacobian matrix have at least
  one zero eigenvalue, then the critical point is non-hyperbolic critical point.
   We summarise the stability criteria of the hyperbolic critical point
   which can be determined from the sign of eigenvalues of $A$ as follows:
   \begin{itemize}
   \item If all the eigenvalues have negative real part then the critical point is stable~(\textit{i.e.}, late time attractor).
   \item If all the eigenvalues have positive real part then the critical point is unstable~(\textit{i.e.}, past time attractor).
   \item If some eigenvalues of the Jacobian matrix have positive real part and other have negative real part then
   the critical point is saddle.
   \end{itemize}

   The above linear stability analysis cannot be used for non-hyperbolic points, so some other techniques such as
   Centre Manifold Theory, Lyapunov's functions \cite{Boehmer:2014vea,wigginsbook} and numerically perturbation of
   solutions around critical points \cite{strogatz} may be employed to study the stability of such points. The method of
   perturbations of solutions around critical points is quite standard and has been extensively used in recent cosmological
   studies \cite{nroy1,nroy,nrn1} and so this technique will be used in our present work.

   In order to solve the above ASODE (\ref{13}) we need to consider specific potential $V(\phi)$
   and coupling function $\mu(\phi)$. Regarding $V(\phi)$, the general popular choice in literature
   is to assume an exponential potential. This potential has the advantage that the phase space remains
   four dimensional and field equations can be studied easily qualitatively using dynamical systems tools.
   In order to consider $V(\phi)$ beyond the constant and exponential potential, one new variable $v\equiv\frac{1}{V}~\frac{dV}{d\phi}$
   has to be introduced \cite{scc,rj,pj}. Further, exponential potential in case of phantom field in standard cosmology
   admits late time attractor but in the RS2 model yields saddle point \cite{GarciaSalcedo:2010wa}. In this paper
   we shall show that it is possible to get late time attractor under this potential in RS2 model when gravity is coupled
   to scalar field. As exponential potentials are known to be significant in dynamical
   investigations \cite{Copeland:1997,gleon,hjj}, we assume $V(\phi)=V_0 e^{-\alpha \phi}$. Concerning coupling function $\mu(\phi)$,
   the popular choice in literature are
   exponential and power-law forms \cite{Mahata:2013oza,gleonn}. In what follows as in
  \cite{Mahata:2013oza}, in order to remain general, we choose two coupling functions (exponential and power-law forms)
  corresponding to same the exponential potential and the
  corresponding analysis will be done separately in following two
  subsections.

\subsection{\bf Scenario 1 : Exponential potential and exponential coupling function}
Here we choose

\begin{equation}\label{22}
V(\phi)=V_0 e^{-\alpha \phi}  \hspace{1.2cm} \textrm{and} \hspace{1.2cm}    \mu(\phi)=\mu_0 e^{\nu \phi}
\end{equation}

In this scenario, the autonomous system (\ref{13}) simplifies to
$$x'=-3x+\alpha_0~y^2 \mathrm{e}^{-\frac{b}{s}}+\frac{3}{2}x\Big(\frac{2-z}{z}\Big)(z+x^2-y^2)$$

$$y'=-\alpha_0~xy \mathrm{e}^{-\frac{b}{s}}+\frac{3}{2}y\Big(\frac{2-z}{z}\Big)(z+x^2-y^2)$$

$$z'= 3(1-z)(z+x^2-y^2)$$
\begin{equation}\label{23}
s'=-\frac{xs^2~\mathrm{e}^{-\frac{b}{s}}}{\sqrt{\mu_0}}
\end{equation}
where $\alpha_0=\alpha~\sqrt{\frac{3}{2\mu_0}}$ and $b=\sqrt{\frac{3}{2}}~\nu$\\
The critical points of the present autonomous system are:\\
$C_{1}(0,0,0,0)$~,~$C_{2}(0,0,1,0)$~,~$C_{3}^{\pm}(\pm 1,0,1,0)$~,~$C_4(0,1,1,s)$~,~$C_5(0,\sqrt{z},z,0)$~,
~$C_6(0,0,0,s)$~,~$C_7(0,0,1,s)$\\
\vspace{2cm}
\begin{center}
Table 1.~Scenario $1$~:~Exponential potential and exponential coupling function. The critical points
for ASODE  (\ref{23}) and values of the relevant parameters\\
\begin{tabular}{c c c c c c c c c}
\hline
\hline
Critical point & x~~~ & y~~~ & z~~~ & s~~~~~ & Existence~~~ & ${\Omega}_{\phi}~~~$ & ${\Omega}_{m}~~~$ & ${\omega}_{\phi}$  \\
\hline\\
$C_1$ & 0~~~ & 0~~~ & 0~~~ & 0~~~~~ & Always~~~ & 0~~~ & 0~~~ & Undefined   \\\\
$C_2$ & 0~~~ & 0~~~ & 1~~~ & 0~~~~~ & ,,~~~ & 0~~~ & 1~~~ & Undefined  \\\\
$C_3^{\pm}$ & $\pm 1$~~~ & 0~~~ & 1~~~ & 0~~~~~ & ,,~~~ & 1~~~ & 0~~~ & 1 \\\\
$C_4$ & 0~~~ & 1~~~ & 1~~~ & 0~~~~~ & ,,~~~ & 1~~~ & 0~~~ & -1 \\\\
$C_5$ & 0~~~ & $\sqrt{z}$~~~ & $z\in ]0,1]$~~~ & 0~~~~~ & ,,~~~ & z~~~ & 0~~~ &-1 \\\\
$C_6$ & 0~~~ & 0~~~ & 0~~~ & s~~~~~ & ,,~~~ & 0~~~ &0~~~ &Undefined \\\\
$C_7$ & 0~~~ & 0~~~ & 1~~~ & s~~~~~ & ,,~~~ & 0~~~ &1~~~ & Undefined \\\\
\hline
\hline
\end{tabular}
\end{center}
\vspace{1cm}
\begin{center}
Table 2.~Scenario $1$~:~Exponential potential and exponential coupling
function. Eigenvalues of the linearised matrix for the  critical
points ASODE (\ref{23}) and corresponding values of dimension of
Stable manifold and deceleration parameters.

\begin{tabular}{c c c c c c c}
\hline
\hline
Critical point & $\lambda_1~~~$ & $\lambda_2~~~$ & $\lambda_3~~~$ & $\lambda_4~~~~~$ &  Stable manifold~~~ & q\\
\hline\\
$C_1$ & $0~~$ & $3~~~$ & $3~~~$ & $0~~~~~$ & No stable manifold~~~ & $2$ \\\\
$C_2$ & $-\frac{3}{2}~~~$ & $\frac{3}{2}~~~$ & $-3~~~$ & $0~~~~~$ & 2D~~~ & $\frac{1}{2}$\\\\
$C_3^{\pm}$ & $3~~$ & $3~~~$ & $-6~~~$ & $0~~~~~$ & 1D~~~ & 2\\\\
$C_4$ & $-3~~$ & $-3~~~$ & $0~~~$ & $0~~~~~$ & 2D~~~ & -1\\\\
$C_5$ & $-3~~$ & $-3~~~$ & $0~~~$ & $0~~~~~$ & 2D~~~ & -1\\\\
$C_6$ & $0~$ & $3~~~$ & $3~~$ & $0~~~~~$ & No stable manifold~~~ &2 \\\\
$C_7$ & $-\frac{3}{2}~~~$ & $\frac{3}{2}~~~$ & $-3~~~$ & $0~~~~~$ & 2D~~~ & $\frac{1}{2}$\\\\

\hline
\hline
\end{tabular}
\end{center}
\subsubsection{\bf Critical points and their properties for Scenario $1$:}

 The critical points of the above ASODE (\ref{23}) and the eigenvalues of the corresponding
 Jacobian matrix are listed in Table $1$ and Table $2$ respectively. Here we see that all the critical points are
 non hyperbolic in nature. So linear stability theory cannot be used
 to study the stability of the system. But using centre manifold theorem we can find dimension of stable
 manifold in this case \cite{wigginsbook}. It may be noted that the five dimensional effects are presented
  when $z\ne 1$ so $C_1$~,~$C_5$~and~$C_6$ only reflects the brane effects.

\begin{figure}[htb]
\begin{center}
\includegraphics[width=9cm,height=7cm]{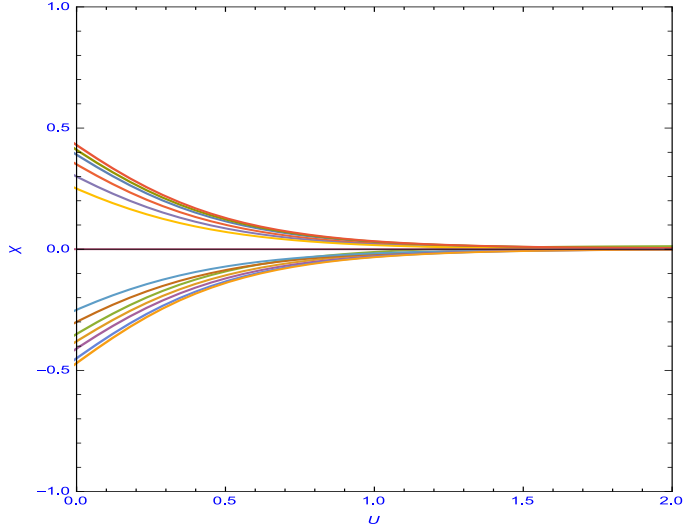}
\vspace{0.3cm}
\caption{The plot of $X$ against $U$ for Scenario $1$ near the critical point~$C_4$}.\label{FIG.1}
\end{center}
\end{figure}

\begin{figure}[htb]
\begin{center}
\includegraphics[width=9cm,height=7cm]{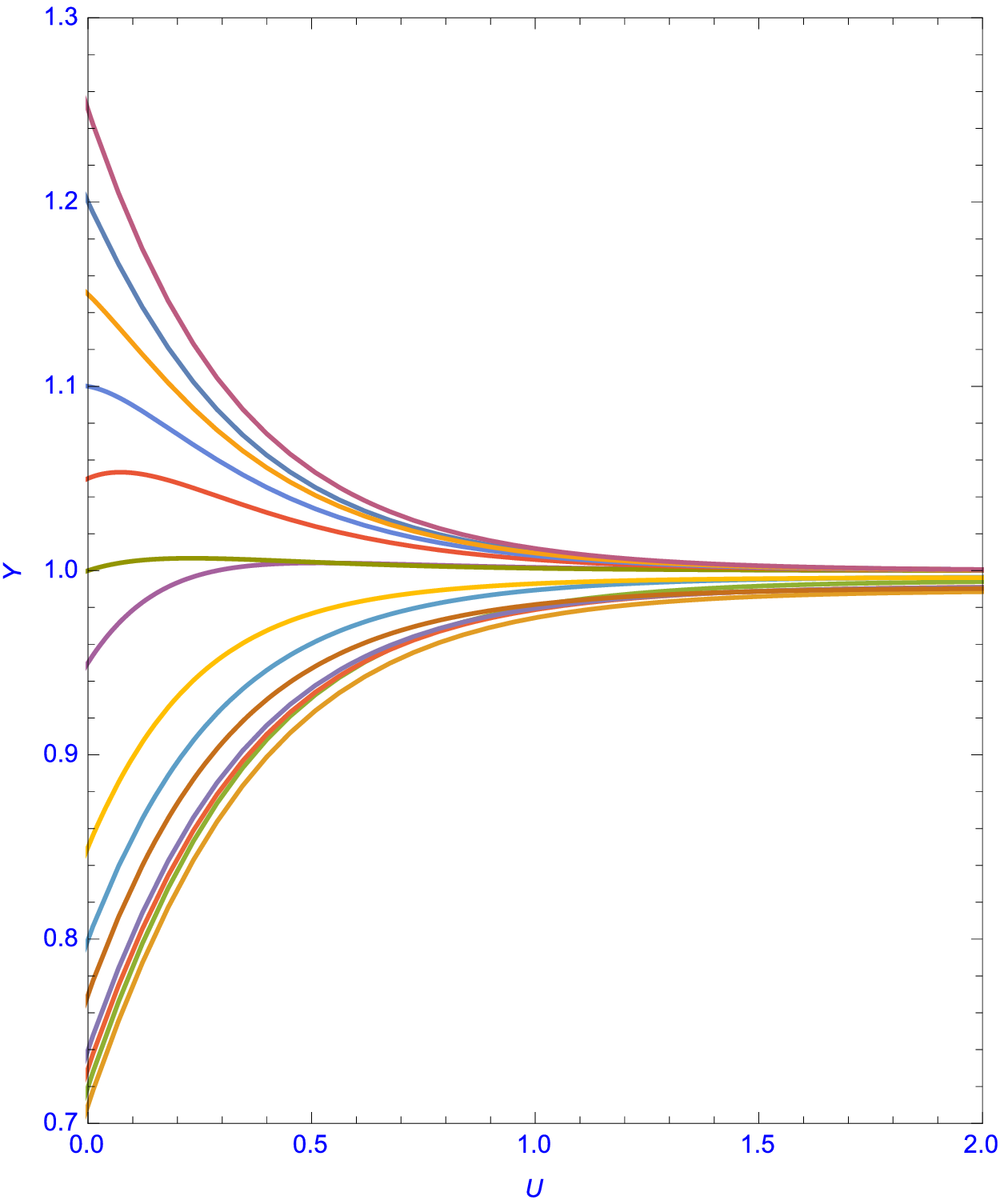}
\vspace{0.3cm}
\caption{The plot of $Y$ against $U$ for Scenario $1$ near the critical point~$C_4$}.\label{FIG.2}
\end{center}
\end{figure}

\begin{figure}[htb]
\begin{center}
\includegraphics[width=9cm,height=7cm]{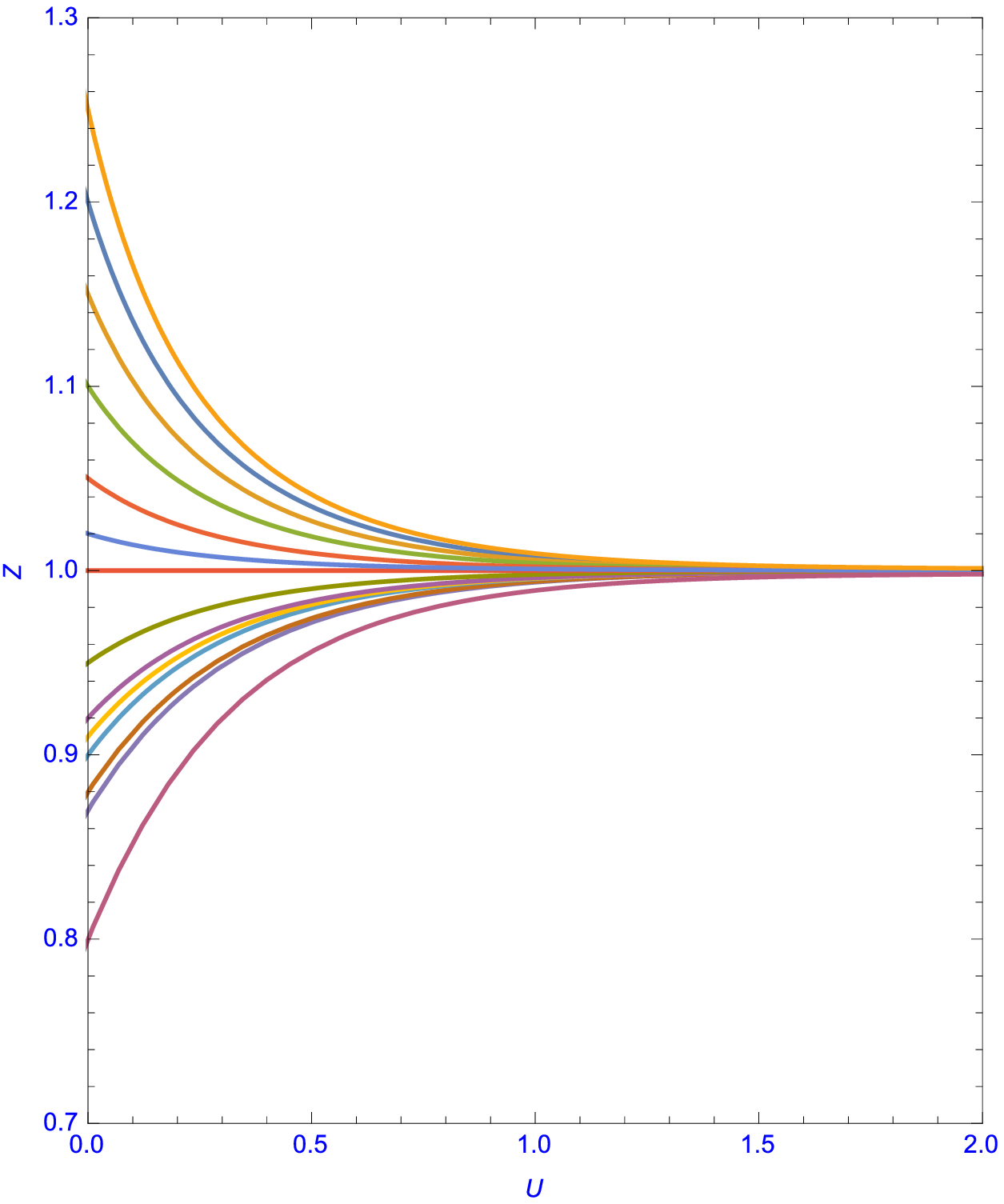}
\vspace{0.3cm}
\caption{The plot of $Z$ against $U$ for Scenario $1$ near the critical point~$C_4$}.\label{FIG.3}
\end{center}
\end{figure}

\begin{figure}[htb]
\begin{center}
\includegraphics[width=9.3cm,height=7.2cm]{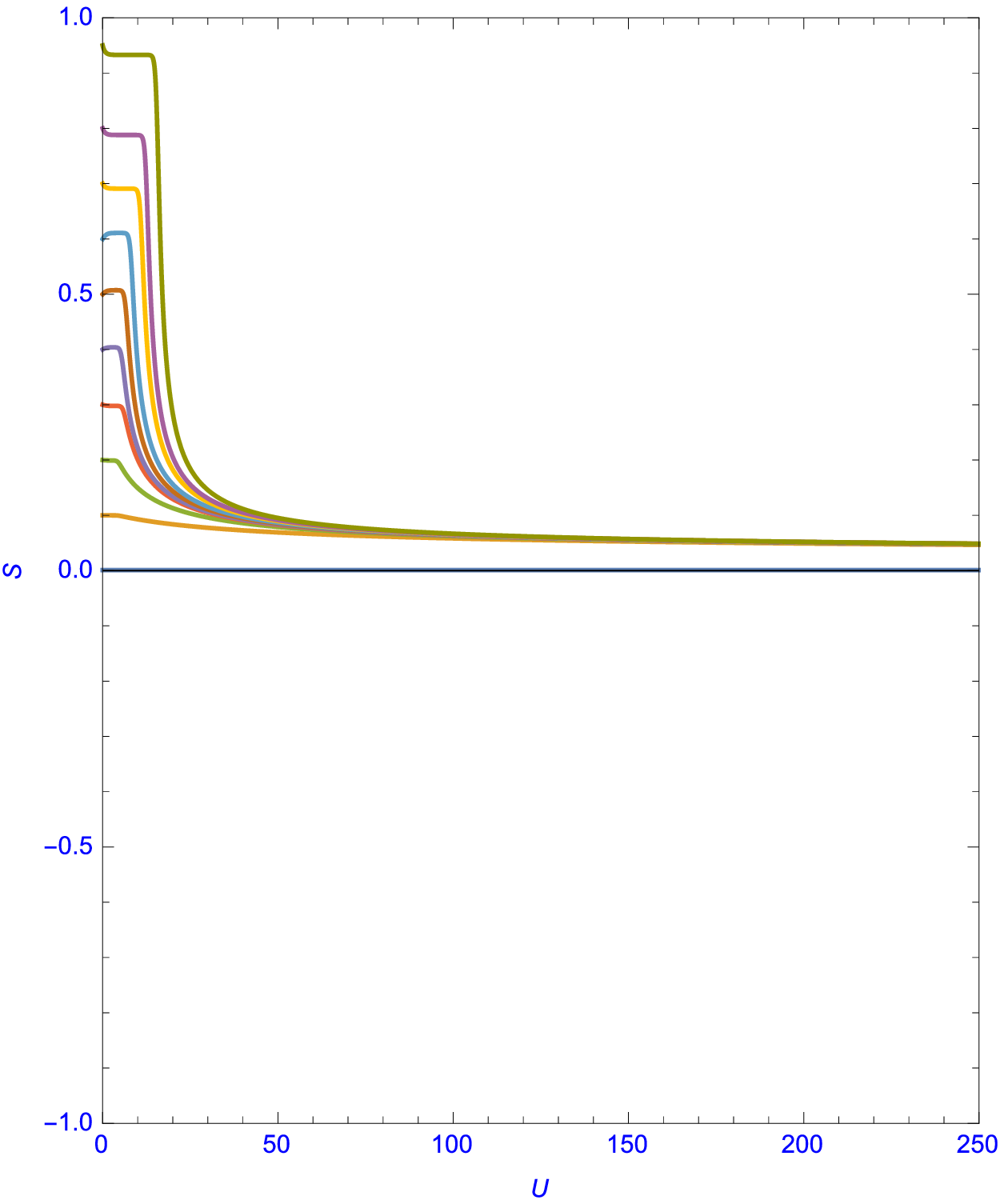}
\vspace{0.3cm}
\caption{The plot of $S$ against $U$ for Scenario $1$ near the critical point~$C_4$}.\label{FIG.4}
\end{center}
\end{figure}

\begin{figure}[htb]
\begin{center}
\includegraphics[width=9.2cm,height=7.2cm]{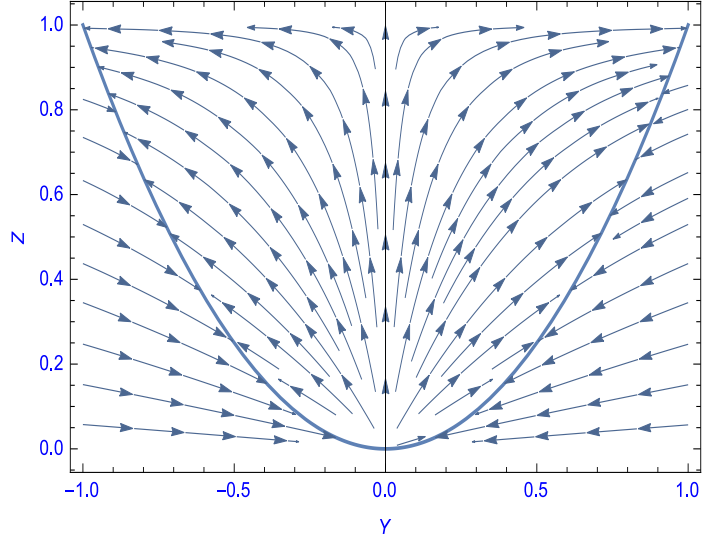}
\vspace{0.3cm}
\caption{Phase portrait of Scenario $1$ near the critical point~$C_5$}.\label{FIG.5}
\end{center}
\end{figure}

  We note that the point $C_1=(0,0,0,0)$ is a critical point of the ASODE (\ref{23}).
 But at this
  point
 the first two equations of ASODE (\ref{23}) are undefined.
 To study the nature of the critical point $C_1$, we can apply a
 secure approach  as reported in \cite{GarciaSalcedo:2010wa}  and expand ASODE (\ref{23}) in
the neighbourhood of  $C_1=(0,0,0,0)$ by putting  $ x\rightarrow 0+\epsilon_x~,~y\rightarrow 0+\epsilon_y~,~z\rightarrow 0+\epsilon_z~$
 and $~s\rightarrow 0+\epsilon_s $ in the above ASODE (\ref{23}).
 If we neglect higher order terms, then the ASODE (\ref{23}) reduces to following system
$\epsilon'_x=0~,~\epsilon'_y=3\epsilon_y~,~\epsilon'_z=3\epsilon_z$~and~$\epsilon'_s=0$.
So, the point $C_1$ is non hyperbolic critical  point. We can
only say in this case is that  the point $(0,0,0,0)$ is unstable as
perturbations will grow exponentially. So this critical point will behave as past time attractor and is
characterised by vanishing matter content and we have empty
(Misner-RS) universe. In braneworld, existence of empty universe
is a distinctive feature. Furthermore, we have a decelerating solution with $(q=2)$~(limiting case) at this point. The behaviour of
$C_6$ is same as $C_1$ and $C_3$ is not interesting from physical point of view. In what follows we summarise
 the remaining critical points of scenario $1$ and their properties.
 \begin{itemize}
 \item As seen from Tables $1$ and $2$, the saddle critical points $C_2$ and $C_7$ indicate that universe is completely dominated by
 dark matter $(\Omega_m=1)$ and expansion of the universe around these points is decelerated
  $(q=\frac{1}{2})$ having 2D stable manifolds.
 These points are saddle in nature.

 \item At the critical point $C_4$, we see that the universe is completely dominated by
 the scalar field, at this point expansion of the universe
 is accelerated $(q=-1)$ having two negative eigenvalues and two zero eigenvalues.
 Since this point is non-hyperbolic, we numerically perturbed the solutions near this critical point and
 check the stability. To this end we plot the projections of perturbations on $x,y,z$ and $s$ axes separately in
  FIG.(\ref{FIG.1}) to  FIG.(\ref{FIG.4}). From  FIG.(\ref{FIG.1}) and  FIG.(\ref{FIG.4}) we see that the projection of
  perturbations solutions asymptotically approach $x=0$ and $s=0$
 respectively as $U$ approaches to infinity. Similarly, from FIG.(\ref{FIG.2})
 and FIG.(\ref{FIG.3}), $y$ and $z$ approach to $1$ as $U$ approaches to
 infinity, so that the whole system asymptotically approach the critical point
 $C_4=(0,1,1,0)$ as $U$ approaches to infinity. From this
 perturbation plots we can conclude that the point $C_4=(0,1,1,0)$ is a
  stable critical point of the ASODE (\ref{23}).

 \item  For the critical point $C_5$, we have $(x,y,z,s)=(0,\sqrt{z},z,0)$. This implies $z=y^2~(x=0~,~s=0)$,
 so that this point is a curves of critical
 points. If we compare $C_4$ and $C_5$, we see that $C_4$ is a particular case of $C_5$. At this point $\rho_m=0$ and $\rho_T=V$, so
 that the above Eq.(\ref{5}) reduces to \\
$3H^2=V~\Big(1+\frac{V}{2\lambda}\Big)$. It represents the slow-roll Friedmann equation  where the Hubble
parameter is related to the potential of the inflation field
having 2D stable manifolds. When  potential is much  larger than
the brane tension $V>>\lambda$~$\Rightarrow
H_{RS}=\frac{V}{\sqrt{6\lambda}}$, so that at early time expansion
rate $H_{RS}$ get enhanced with respect to the $H_{GR}$ ( GR rate
of expansion) in the RS2 model and we have
$\frac{H_{RS}}{H_{GR}}=\sqrt{\frac{V}{2\lambda}}$. Thus in this
way brane effects fuel early inflation in RS2 model. FIG.(\ref{FIG.5}) shows that the nature of curves of critical points of $C_5$
\end{itemize}

\subsection{\bf Scenario $2$ : Exponential potential and power law coupling parameter}
\noindent Here we choose
\begin{equation}\label{24}
V(\phi)=V_0 e^{-\alpha \phi}  \hspace{1.2cm} \textrm{and}\hspace{1.2cm}    \mu(\phi)=\mu_0  \phi^{n}
\end{equation}\\
\noindent where $V_0$, $\mu_0$ and  $n$ are constants.\\
Then the autonomous system~(\ref{13})~becomes

$$x'=-3x+\beta_2~y^2s^{\frac{n}{2}}+\frac{3}{2}x\Big(\frac{2-z}{z}\Big)(z+x^2-y^2)$$
$$y'=-\beta_2~xys^{\frac{n}{2}}+\frac{3}{2}y\Big(\frac{2-z}{z}\Big)(z+x^2-y^2)$$
$$ z'= 3(1-z)(z+x^2-y^2)$$
\begin{equation}\label{25}
s'=-\mu_2~xs^{2+\frac{n}{2}}
\end{equation}
\noindent where $\beta_2=\alpha_06^{-\frac{n}{4}}$ and $\mu_2=\frac{6^{-\frac{n}{4}}}{\sqrt{\mu_0}}$.
The critical points of the present autonomous system are $D_1(0,0,0,0)$~,~$D_{2}(0,0,1,0)$~,
~$D_{3}^{\pm}(\pm 1,0,1,0)$~,~$D_4(0,1,1,s)$~,~$D_5(0,\sqrt{z},z,0)$~,~$D_6(0,0,0,s)$~,~$D_7(0,0,1,s)$.\\
\vspace{2cm}
\begin{center}
Table 3.~Scenario $2$~:~Exponential potential and power law coupling function. The critical points
for ASODE  (\ref{25}) and values of the relevant parameters\\
\begin{tabular}{c c c c c c c c c}
\hline
\hline
Critical point & x~~~ & y~~~ & z~~~ & s~~~~~ & Existence~~~ & ${\Omega}_{\phi}~~~$ & ${\Omega}_{m}~~~$ & ${\omega}_{\phi}$  \\
\hline\\
$D_1$ & 0~~~ & 0~~~ & 0~~~ & 0~~~~~ & Always~~~ & 0~~~ & 0~~~ & Undefined   \\\\
$D_2$ & 0~~~ & 0~~~ & 1~~~ & 0~~~~~ & ,,~~~ & 0~~~ & 1~~~ & Undefined  \\\\
$D_3^{\pm}$ & $\pm 1$~~~ & 0~~~ & 1~~~ & 0~~~~~ & ,,~~~ & 1~~~ & 0~~~ & 1 \\\\
$D_4$ & 0~~~ & 1~~~ & 1~~~ & 0~~~~~ & ,,~~~ & 1~~~ & 0~~~ & -1 \\\\
$D_5$ & 0~~~ & $\sqrt{z}$~~~ & $z\in ]0,1]$~~~ & 0~~~~~ & ,,~~~ & z~~~ & 0~~~ &-1 \\\\
$D_6$ & 0~~~ & 0~~~ & 0~~~ & s~~~~~ & ,,~~~ & 0~~~ &0~~~ &Undefined \\\\
$D_7$ & 0~~~ & 0~~~ & 1~~~ & s~~~~~ & ,,~~~ & 0~~~ &1~~~ & Undefined \\\\
\hline
\hline
\end{tabular}
\end{center}
\vspace{1cm}
\begin{center}
Table 4.~Scenario $2$~:~Exponential potential and power law coupling
function. Eigenvalues of the linearised matrix for the  critical
points ASODE (\ref{25}) and corresponding values of dimension of
Stable manifold and deceleration parameters.

\begin{tabular}{c c c c c c c}
\hline
\hline
Critical point & $\lambda_1~~~$ & $\lambda_2~~~$ & $\lambda_3~~~$ & $\lambda_4~~~~~$ &  Stable manifold~~~ & q\\
\hline\\
$D_1$ & $0~~$ & $3~~~$ & $3~~~$ & $0~~~~~$ & No stable manifold~~~ & $2$ \\\\
$D_2$ & $-\frac{3}{2}~~~$ & $\frac{3}{2}~~~$ & $-3~~~$ & $0~~~~~$ & 2D~~~ & $\frac{1}{2}$\\\\
$D_3^{\pm}$ & $3~~$ & $3~~~$ & $-6~~~$ & $0~~~~~$ & 1D~~~ & 2\\\\
$D_4$ & $-3~~$ & $-3~~~$ & $0~~~$ & $0~~~~~$ & 2D~~~ & -1\\\\
$D_5$ & $-3~~$ & $-3~~~$ & $0~~~$ & $0~~~~~$ & 2D~~~ & -1\\\\
$D_6$ & $0~$ & $3~~~$ & $3~~$ & $0~~~~~$ & No stable manifold~~~ &2 \\\\
$D_7$ & $-\frac{3}{2}~~~$ & $\frac{3}{2}~~~$ & $-3~~~$ & $0~~~~~$ & 2D~~~ & $\frac{1}{2}$\\\\

\hline
\hline
\end{tabular}
\end{center}
\subsubsection{\bf Critical points and their properties for Scenario $2$:}

 As before all the critical points are non
 hyperbolic in nature. Tables $3$ and $4$ shows the relevant parameters, eigenvalues of the linearised matrix
 along with the corresponding values of dimension of stable manifold and deceleration parameter at the
 critical points. We apply the same technique to study the nature of
 these points and the cosmological implications at each critical points are also same with Scenario $1$.

\section{Critical points of 3D autonomous system}\label{IV}

 As all the critical points are non hyperbolic in nature in above 4D ASODE and so present linear stability analysis can not be used
 to study such system consistently. However, if the potential function $V(\phi)$ and the coupling function $\mu(\phi)$ are
 chosen such that $\beta(\phi)$ is either constant or a function of $x$,$y$,$z$ only, then the above ASODE (\ref{13}) can be reduced to 3D.
 In what follows we shall analyse ASODE (\ref{13}) for some choices of $V(\phi)$  and $\mu(\phi)$ such that\\\\
 $~~~(i)~~\beta(\phi)$=\rm{constant}~~and~~$~~~(ii)~~\beta(\phi)=-\frac{1}{2x}$
 \subsection{\bf When $\beta$ is a constant.}

  As in Ref \cite{Mahata:2013oza}, in this case the coupling function can be either constant or in power law form while potential function
  can be in exponential and power law form.

\noindent Then the 4D ASODE (\ref{13}) reduces to the 3D autonomous system as
$$x'=-3x-\beta~y^2+\frac{3}{2}x\Big(\frac{2-z}{z}\Big)(z+x^2-y^2)$$

$$y'=\beta~xy+\frac{3}{2}y\Big(\frac{2-z}{z}\Big)(z+x^2-y^2)$$
\begin{equation}\label{26}
z'= 3(1-z)(z+x^2-y^2)
\end{equation}
The critical points of the present autonomous systems are:\\
$E_1(0,0,0)$~,~$E_2(0,0,1)$~,~$E_3^\pm(\pm 1,0,1)$~,~$E_4\Big(-\frac{\beta}{3},\sqrt{1-\frac{\beta^{2}}{9}},1\Big)$~,~$E_5\Big(-\frac{3}{2\beta},\frac{3}{2\beta},1\Big)$.
\begin{center}
Table 5.~The critical points
for ASODE  (\ref{26}) and values of the relevant parameters\\
\begin{tabular}{c c c c c c c c}
\hline
\hline
Critical point & x~~~ & y~~~ & z~~~ & Existence~~~ & ${\Omega}_{\phi}~~~$ & ${\Omega}_{m}~~~$ & ${\omega}_{\phi}$  \\
\hline\\
$E_1$ & 0~~~ & 0~~~ & 0~~~ & Always~~~ & 0~~~ & 0~~~ & Undefined   \\\\
$E_2$ & 0~~~ & 0~~~ & 1~~~ & ,,~~~ & 0~~~ & 1~~~ & Undefined   \\\\
$E_3^{\pm}$ & $\pm$1~~~ & 0~~~ & 1~~~ & ,,~~~ & 1~~~ & 0~~~ & 1 \\\\
$E_4$ & $-\frac{\beta}{3}$~~~ &$\sqrt{1- \frac{\beta^{2}}{9}}$~~~ & 1~~~ & $\beta^{2}<9$~~~ & 1~~~ & 0~~~ & $-1+\frac{2\beta^{2}}{9}$\\\\
$E_5$ & $-\frac{3}{2\beta}$~~~ & $\frac{3}{2\beta}$~~~ & 1~~~ & $\beta^{2}>\frac{9}{2}~~~$ & $\frac{3}{\beta}$~~~ & $1-\frac{3}{\beta}$~~~ & $0$\\\\

\hline
\hline
\end{tabular}
\end{center}
\vspace{1cm}
\bigskip
\begin{center}
Table 6.~Eigenvalues of the linearised matrix for the critical
points of ASODE (\ref{26}) and corresponding values of dimension of
Stable manifold and deceleration parameters.
 We have defined :
$A\equiv \frac{\sqrt{36-7\beta^{2}}}{\beta}$.\\\

\begin{tabular}{c c c c c c}
\hline
\hline
Critical point & $\lambda_1~~~$ & $\lambda_2~~~$ & $\lambda_3~~~$ &  Stable manifold~~~ & q\\
\hline\\
$E_1$ & $0~~~$ & $3~~~$ & $3~~~$ & No stable manifold~~~ & $2$ \\\\
$E_2$ & $-\frac{3}{2}~~~$ & $\frac{3}{2}~~~$ & $-3~~~$ & 2D~~~ & $\frac{1}{2}$ \\\\
$E_3^\pm$ & $3~~~$ & $3\pm\beta~~~$ & $-6~~~$ & 2D if $3<\beta~~~$ & 2\\\\
$E_4$ & $-\frac{1}{3}(9-2\beta^{2})~~~$ & $-\frac{1}{3}(9-\beta^{2})~~~$ & $-\frac{2\beta^{2}}{3}~~~$ & ~~3D
if $\beta^{2}<\frac{9}{2}$~~~ & $\frac{1}{3}(\beta^{2}-3)$\\\\
$E_5$ & $-\frac{3}{4}(1+A)~~~$ & $-\frac{3}{4}(1-A)~~~$ & $-3~~~$ & ~~3D~if $A<1~~~$ & $\frac{1}{2}$ \\\\
\hline
\hline
\end{tabular}
\end{center}
\vspace{0.3cm}
\subsubsection {\bf Critical points and their properties when
$\beta$ is constant }
\begin{figure}[htb]
\begin{center}
\includegraphics[width=7cm,height=6cm]{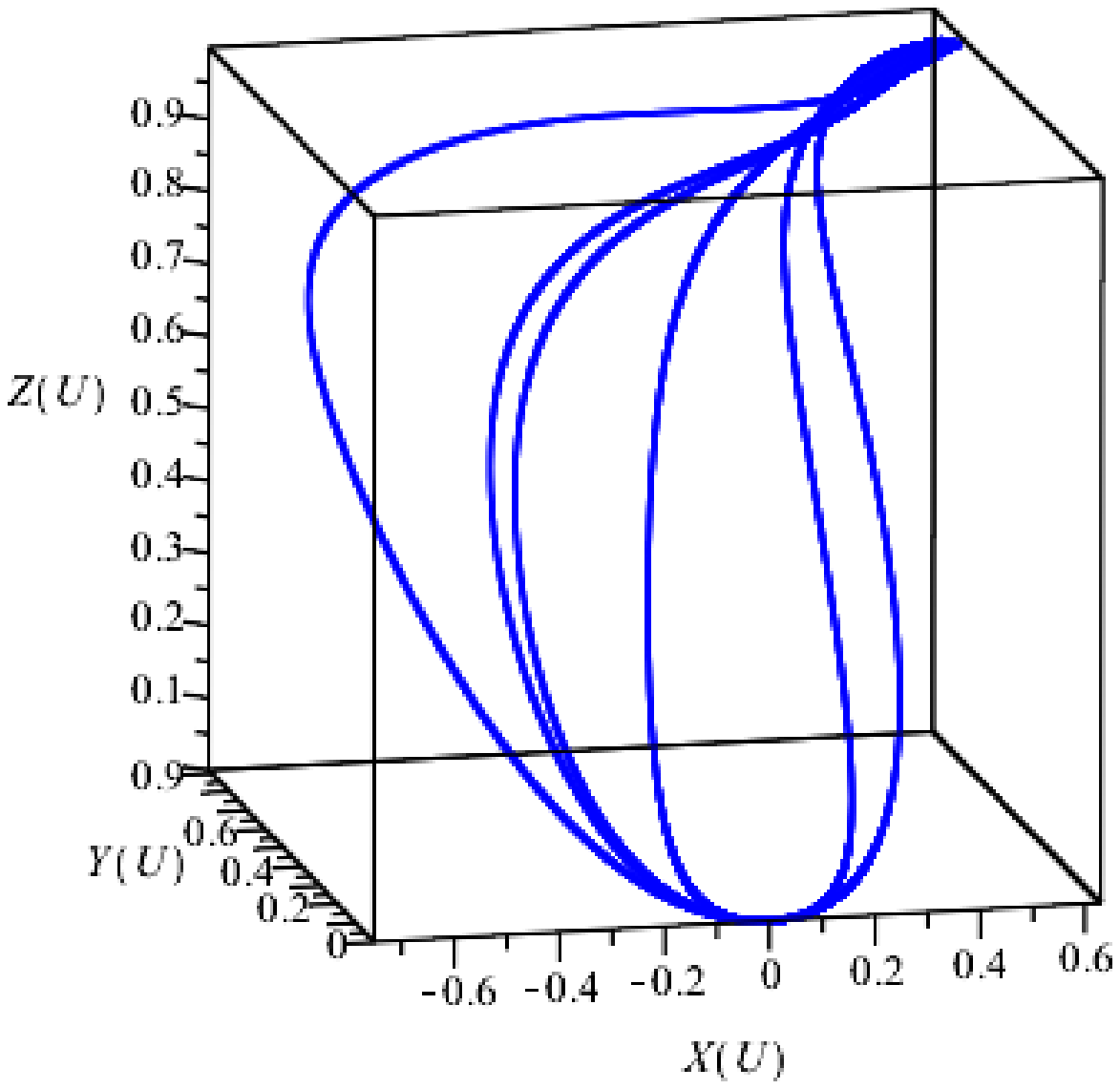}
\includegraphics[width=7cm,height=6cm]{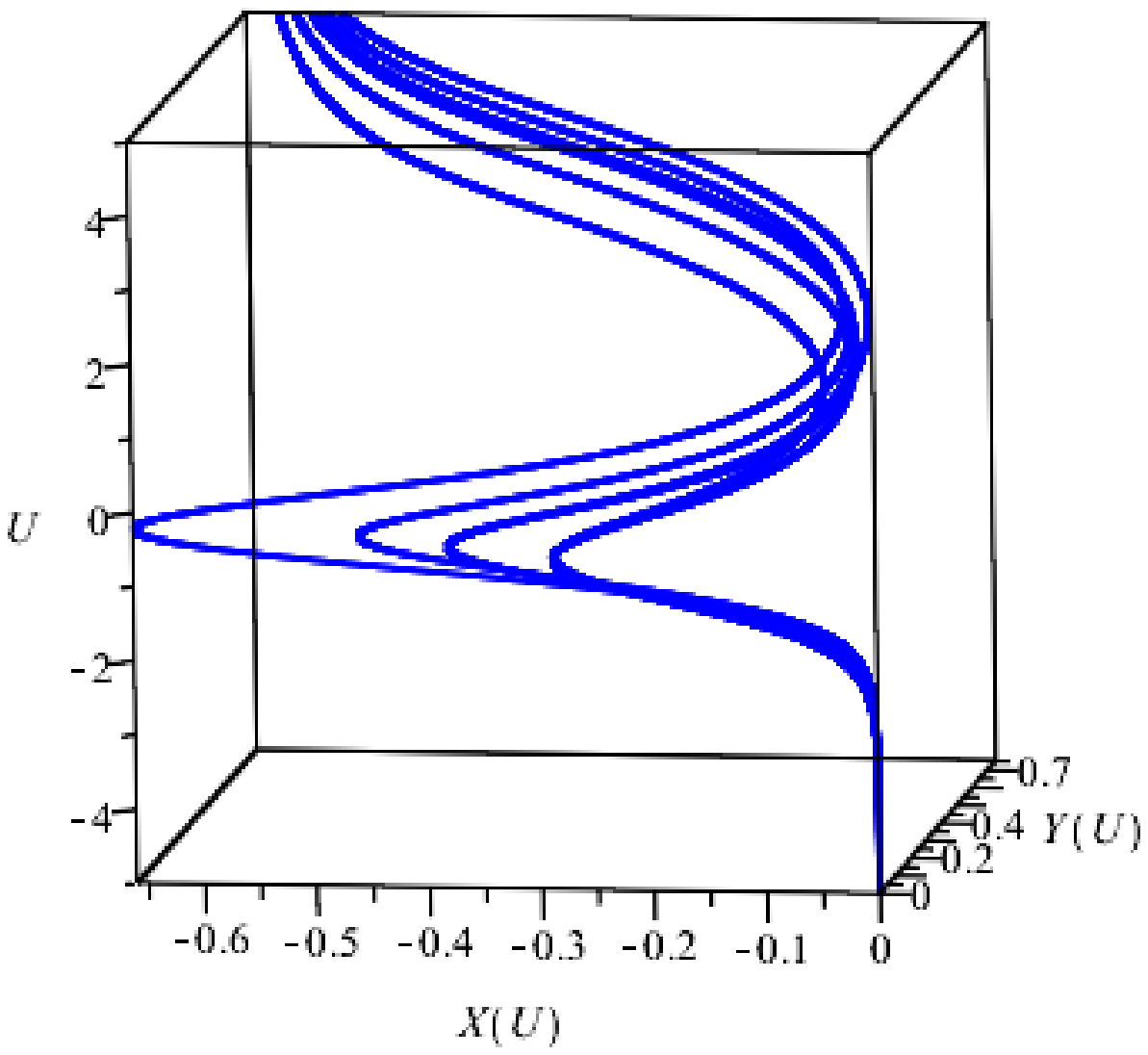}
\end{center}
\vspace{0.02cm}
\bigskip
\caption ~{Trajectories in phase space $(x,y,z)$ for different sets of initial conditions with
 different values of  parameter
$\beta=2$ (left) and the flux in time $U=\ln a $ (right)~from these figures we see that
  the scalar field dominated solution is the late-time attractor.}
\label{FIG.6}
\end{figure}

\begin{figure}[htb]
\begin{center}
\includegraphics[width=9cm,height=7cm]{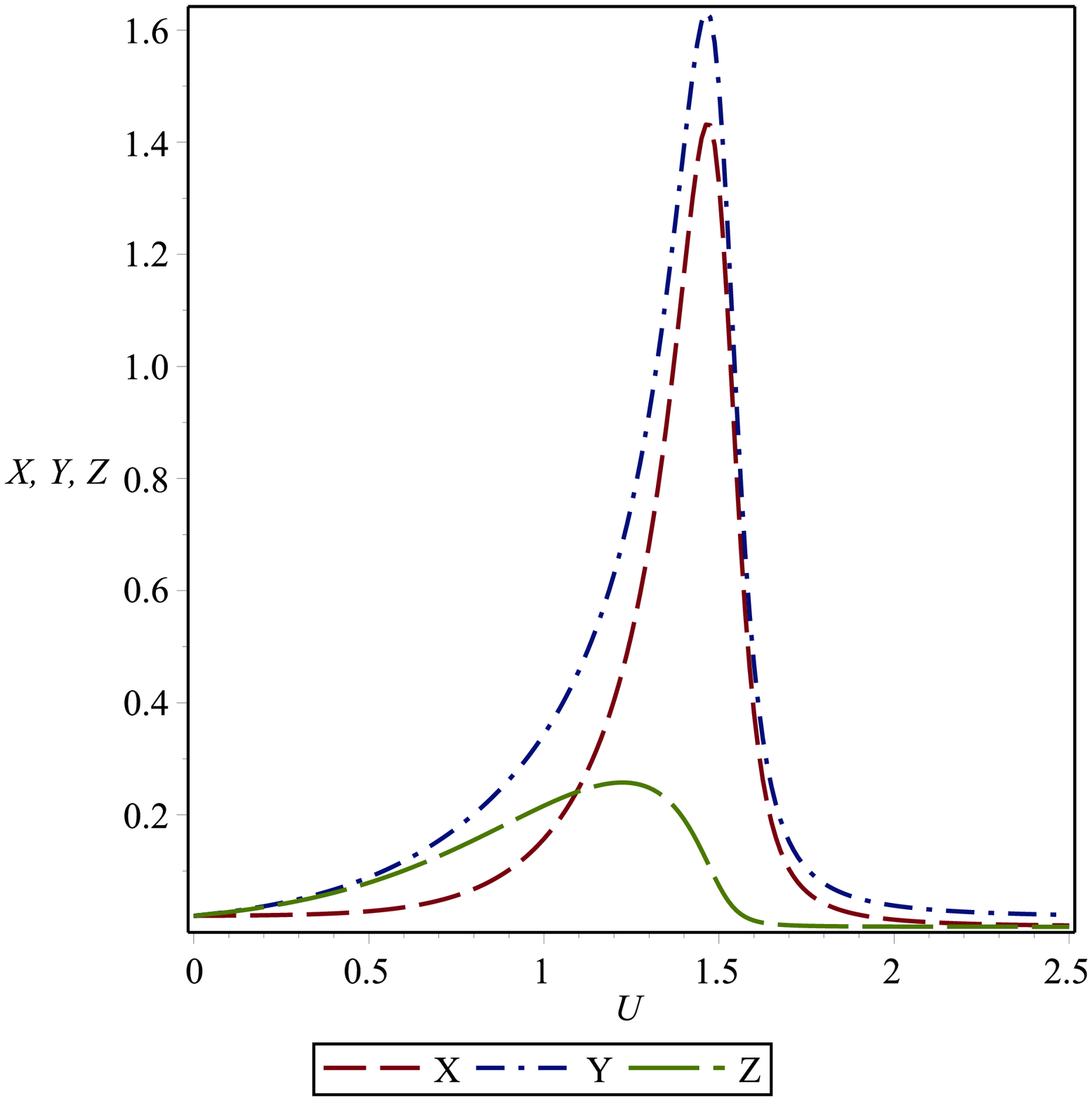}
\end{center}
\vspace{0.3cm}
\bigskip
\caption ~{The dimensionless auxiliary variables are plotted
against e-folding time $(U=\ln a)$ for $\beta=-5$} \label{FIG.7}
\end{figure}

Here the point $E_1=(0,0,0)$ is a critical point of the ASODE (\ref{26}). The evolution of the ASODE (\ref{26})
are shown in FIG.(\ref{FIG.6}) and FIG.(\ref{FIG.7}) numerically. From FIG.(\ref{FIG.6}), we see
that this point is the past attractor of the RS cosmological model
and we have a decelerating solution with $(q=2)$~(limiting case).
This is the only
 critical point with $z \ne 1$ and we get five dimensional
 behaviour. The remaining critical points and their properties which are associated with four
dimensional behaviour are summarised as follows.

\begin{figure}[htb]
\begin{center}
\includegraphics[width=8cm,height=7cm]{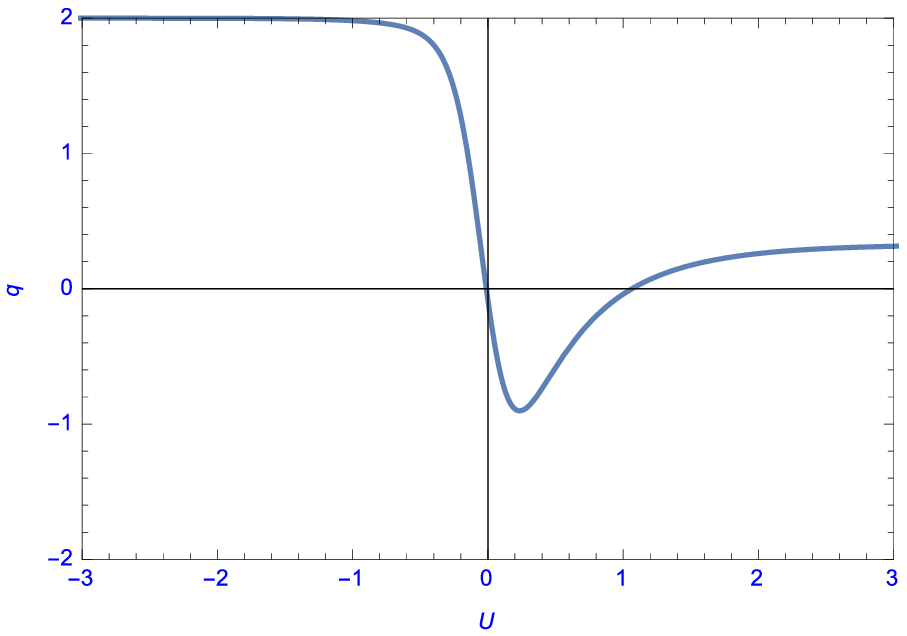}
\vspace{0.3cm}
\caption{The plot of $q$  against $U$ for $\beta=2$}.\label{FIG.8}
\end{center}
\end{figure}

\begin{itemize}
 \item For the critical point $E_2$, the universe is
 completely dominated by dark matter $(\Omega_m=1)$ and the EoS parameter $\omega_\phi$ is undefined.
 As seen from Table 6 this point is saddle critical point in the phase space.
 At this point the expansion of universe is decelerated $(q=\frac{1}{2})$.

\item The critical points $E_3^\pm$ correspond to solution where the universe
is completely dominated by the kinetic energy of the scalar field $(\Omega_\phi=1)$
with an equation of state parameter $(\omega_\phi=1)$. These critical points are also saddle critical points
 in the phase space  and expansion of the universe is also decelerated at these points $(q=2)$.

\item For the critical point $E_4$, we see that the universe is dominated by the scalar
 field (see Table 5) and we have late time attractor of the ASODE (\ref{26}) and this point will
 be stable in the phase space of the Brane scenario for $\beta^{2}<\frac{9}{2}$. At this point,
 there exists an accelerating phase of the universe for  $\beta^{2}<3$.\\\

\item For $\beta$ constant, the matter scaling solution is
represented by the critical point
 $E_5$ and it is the late time attractor if $A<1$~(for $\frac{9}{2}<\beta^2<\frac{36}{7}$).
 The expansion of the universe around this point is decelerated $(q=\frac{1}{2})$.
  Here the scalar field behaves as dust $(\omega_\phi=0)$.
  This corroborates  earlier analytical investigations where it was shown that in RS-model there will be
   an decelerated expansion after the accelerated expansion \cite{sk}. FIG.(\ref{FIG.8}) shows that
universe starts with a deceleration $(q=2)$ and finally evolves to phase of deceleration $(q=0.30)$.
\end {itemize}
 The points $E_2$, $E_3^\pm$  correspond to the
 critical points $P_1$, $P_{2}^{\pm}$ of \cite{Gonzalez:2008wa} and the cosmological implications are
 same.

\subsection{ \bf When $\beta(\phi)=-\frac{1}{2x}$.}
The form of the potential function in this case is given by\\
\begin{center}
$V=\frac{V_0}{a}$
\end{center}
where $a$ is the scale factor.
Then the autonomous system (\ref{13}) reduces to\\
$$x'=-3x+\frac{y^2}{2x}+\frac{3}{2}x\Big(\frac{2-z}{z}\Big)(z+x^2-y^2)$$

$$y'=-\frac{y}{2}+\frac{3}{2}y\Big(\frac{2-z}{z}\Big)(z+x^2-y^2)$$
\begin{equation}\label{27}
z'= 3(1-z)(z+x^2-y^2)
\end{equation}
The critical points of the above autonomous systems are:\\
$E_6(0,0,0)$~,~$E_7^{\pm}(\pm 1,0,1)$ and $E_8^{\pm}\Big(\pm \frac{1}{\sqrt{6}},\sqrt{\frac{5}{6}},1\Big)$.\\\
\vspace{0.5cm}
\begin{center}
Table 7.~The critical points
for ASODE  (\ref{27}) and values of the relevant parameters\\
\begin{tabular}{c c c c c c c c}
\hline
\hline
Critical point & x~~~ & y~~~ & z~~~ & Existence~~~ & ${\Omega}_{\phi}~~~$ & ${\Omega}_{m}~~~$ & ${\omega}_{\phi}$  \\
\hline\\
$E_6$ & 0~~~ & 0~~~ & 0~~~ & Always~~~ & 0~~~ & 0~~~ & Undefined   \\\\
$E_7^\pm$ & $\pm 1$~~~ & 0~~~ & 1~~~ & ,,~~~ & 1~~~ & 0~~~ & 1   \\\\
$E_8^{\pm}$ & $\pm\frac{1}{\sqrt{6}}$~~~ & $\sqrt{\frac{5}{6}}$~~~ & 1~~~ & ,,~~~ & 1~~~ & 0~~~ & $-\frac{2}{3}$ \\\\
\hline
\hline
\end{tabular}
\end{center}
\hspace{1cm}
\bigskip
\begin{center}
Table 8.~Eigenvalues of the linearised matrix for the critical
points of ASODE (\ref{27}) and corresponding values of dimension of
Stable manifold and deceleration parameters.
 We have defined :
$B\equiv 13+\sqrt{129}$~ and~ $C\equiv 13-\sqrt{129}$.\\\
\begin{tabular}{c c c c c c}
\hline
\hline
Critical point & $\lambda_1~~~$ & $\lambda_2~~~$ & $\lambda_3~~~$ &  Stable manifold~~~ & q\\
\hline\\
$E_6$ & $0~~~$ & 3~~~ & 3 & No stable manifold~~~ & $2$ \\\\
$E_7^\pm$ & $3~~~$ & $\frac{5}{2}~~~$ & $-6~~~$ & 1D~~~ & $2$ \\\\
$E_8^\pm$ & $-\frac{B}{8}~~~$ & $-\frac{C}{8}~~~$ & $-1~~~$ & 3D ~~~ & $-\frac{1}{2}$\\\\
\hline
\hline
\end{tabular}
\end{center}
\subsubsection{\bf Critical points and their properties when $\beta=-\frac{1}{2x}$:}
\begin{figure}[htb]
\begin{center}
\includegraphics[width=7cm,height=6cm]{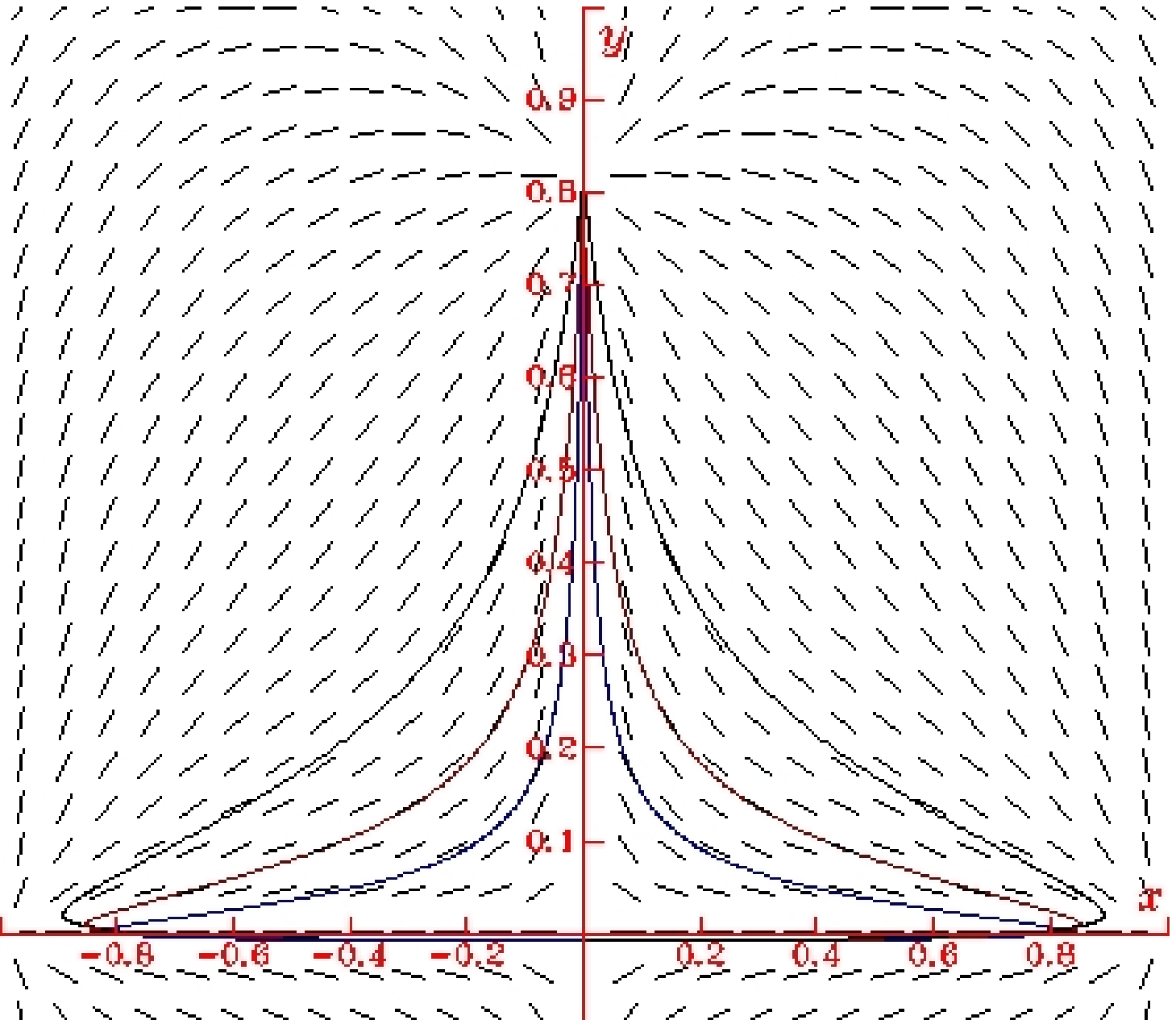}
\vspace{0.3cm}
\bigskip
\caption{Projections of the phase trajectories on to the plane
$(x,y)$ for $\beta=-\frac{1}{2x}$} \label{FIG.9}
\end{center}
\end{figure}

\begin{itemize}
\item The critical point $E_6$ represents the past attractor and the cosmological implications of $E_6$ and $E_1$ are same.

\item At the critical points $E_7^\pm$, the universe is completely dominated by the scalar field $(\Omega_\phi=1)$. As seen from
Table 8 this critical points are saddle in nature. At these points expansion of the universe is also decelerated $(q=2)$.

\item For the critical points $E_8^\pm$, we see that universe is dominated by the scalar field (see Table 7) and we
have late-time attractor of the ASODE (\ref{27}). These points are stable in the phase space and the
expansion of the universe is accelerated around these points $(q=-\frac{1}{2})$.  Projection of the phase
trajectories on to  the $xy$ plane for $\beta=-\frac{1}{2x}$ is given in FIG.(\ref{FIG.9})
\end{itemize}
\section{Classical Stability Of the Model }\label{V}

  In previous sections we have studied local stability of the
   model. In order to predict the final evolution of the universe, it is also desirable
   to study classical stability of the model.

	In cosmological perturbation theory, the speed of sound $(C_s)$ is defined by
	\begin{equation} \label{28}
	C^2_s=\frac{\frac{\partial p}{\partial U}}{\frac{\partial \rho}{\partial U}}
	\end{equation}
	 which  appears as a co-efficient of the $\frac{k^2}{a^2}$ term, where $k$ is the co-moving momentum and $a$ is
 the usual scale factor. The model is said to be classically stable if  $C^2_s$ is positive at local critical points
 $(x,y,z,s)$\cite{Mahata:2013oza,pf}.
 From cosmological point of view only those points which are locally as well as classically stable are interesting.

 \subsection{\bf Scenario 1 : Exponential potential and exponential coupling function}
 In the Scenario 1 we have
\begin {equation}\label{29}
C^2_s=1-\frac{\alpha~{\sqrt{2}}~{\mathrm{e}^{-\frac{b}{s}}}~y^2} {\sqrt{3~\mu_0}~x}
\end {equation}

so for classical stability\\
\begin {equation}\label{30}
\sqrt{3~\mu_0}~x \geq \alpha~{\sqrt{2}}~{\mathrm{e}^{-\frac{b}{s}}}~y^2
\end {equation}

\noindent The local as well as classical stability criteria at critical points are summarised in Table $9$.
It may be noted
 we can not infer about the stability of the model for the critical points
$C_1$~,~$C_2$~,~$C_6$~and~$C_7$ conclusively from above relation
and for that reason we do not include these  points in  Table 9.
From this table we can see that only $C_4$ and $C_5$ are
cosmologically interesting.
\vspace{1cm}
\begin{center}
Table 9.~Scenario 1: Condition for classical stability  at the critical point.\\\
\begin{tabular}{c  c  c  c  l  l l }
\hline
\hline
Critical point & x~~~ & y~~~ & z~~~ & s~~~~~ & Local stability~~~~~ & Classical stability \\
\hline\\
$C_3^\pm$ &$\pm1$~~~ & 0~~~ &  1~~~ & 0~~~~~ &  Unstable(saddle)~~~~~ & Stable\\\\
$C_4$ & 0~~~ & 1~~~ &  1~~~ & 0~~~~~ &  Stable~~~~~ & Stable(limiting) \\\\
$C_5$ & 0~~~  &$\sqrt{z}$~~~ &  z~~~ & 0~~~~~ &  Stable~~~~~ & Stable(limiting) \\\\
\hline
\hline
\end {tabular}
\end{center}

\subsection{\bf Scenario 2 : Exponential potential and power law
coupling parameter}
\noindent In this case, the calculation for $C^2_s$ is obtained as\\
\begin {equation}\label{31}
C^2_s=1-\frac{\alpha~{\sqrt{2}}~{{s}^{\frac{n}{2}}}~y^2} {\sqrt{3~\mu_0}~x}
\end {equation}
so for classical stability\\
\begin {equation}\label{32}
\sqrt{3\mu_0}~x \geq \alpha~{\sqrt{2}}~{{s}^{\frac{n}{2}}}~y^2
\end {equation}

 For this case, local stability and classical stability analysis for the critical points $D_1$-$D_7$ of ASODE
 (\ref{24}) are same as Scenario 1.

\subsection{\bf When $\beta$ is constant and $\beta=-\frac{1}{2x}$}
\noindent In these cases, the calculations for $C^2_s$ are same as with GR case \cite{Mahata:2013oza}.
 We shall only present  conditions for local and classical stability of the
model in Table 10. Here also
 we can not infer about the stability of the model for the critical points
$E_1$~,~$E_2$~and~$E_6$ conclusively from the relation.  From this table
we can see that only $E_4$ and $E_5$ are cosmologically
interesting. \hspace{10cm}
\bigskip
\begin{center}
 Table 10.: Condition for classical stability  at the critical point.\\\
\begin{tabular}{c  c  c  l  l l }
\hline
\hline
Critical point & x~~~ & y~~~ & z~~~ & Local stability~~~~~ & Classical stability \\
\hline\\
$E_3^\pm$ &$\pm1$~~~ & 0~~~ &  1~~~ & Unstable(saddle)~~~~~ & Stable\\\\
$E_4$ & $-\frac{\beta}{3}$~~~ &$\sqrt{1-\frac{\beta^{2}}{9}}$~~~ &  1~~~ & Stable(node)~~~~~ & Stable
if $\frac{9}{2}<\beta^{2}<9$ \\\\
$E_5$ &$-\frac{3}{2\beta}$~~~  &$\frac{3}{2\beta}$~~~ &  1~~~ &  Stable(node)~~~~~ & Stable(limiting) \\\\
$E_7^\pm$ & $\pm 1$~~~& 0 ~~~ & 1~~~ & Unstable(saddle)~~~~~ & Stable\\\\
$E_8^\pm$ &$\pm\frac{1}{\sqrt{6}}$~~~  &$\sqrt{\frac{5}{6}}$~~~ & 1~~~ &  Stable(node)~~~~~ & Unstable \\\\
\hline
\hline
\end {tabular}
\end{center}

\section{Discussion and conclusion}

 A dynamical system analysis of RS2 model of braneworld (when
gravity is coupled to scalar field with a coupling function and
potential) has been analysed in this work. We have taken two coupling
functions (exponential and power law) corresponding to the same
exponential  usual potential. Performing a detailed phase-space
analysis of these two scenarios, we have extracted stable
solutions along with all relevant physical cosmological parameters
($\Omega_\phi$,~$\Omega_m$,~$\omega_\phi$,~$q$),
dimension of stable manifold and eigenvalues of the corresponding
Jacobian matrices in Tables $1-8$. More or less, we get the same results for both scenarios.

Generally the dynamics of RS2 model differs from standard GR based
models. In RS2 model past attractors characterise empty (Misner
RS) universe in contrast to GR based result where the kinetic
dominated solution is the past attractor.

While in 4D analysis we have numerically perturbed the solutions around the non-hyperbolic
critical points, the linear stability analysis is enough to study the critical points in 3D. In both
4D and 3D analysis of critical points we see that it is possible to find past attractor, saddle points and future
attractors under some conditions.  So  it fulfils our wish list and we get a complete
cosmic scenario. This result is to be contrasted with
similar GR based work \cite{Mahata:2013oza} where we do not get
complete cosmic scenario. When $\beta$=\rm constant, we get one
additional future attractor where universe will undergo
decelerated expansion after accelerated expansion. This is a very
interesting result which corroborates earlier results.
Furthermore, we get matter scaling solution which are future
attractor when $\beta$=\rm constant. Thus, the combined effects of
coupling and brane effects gives rich
dynamics in contrast to interaction of dark energy in RS2
braneworld \cite{Biswas:2015zka}. Here past attractor
represents empty universe which is a distinctive feature of
braneworld.

  In addition to local stability, we have also investigated the classical stability of the model.
  Classical stability plays an important role in deciding final state of the universe.
  Only those points which are locally as well as classically stable are interesting from
  cosmological point of view. While in 4D analysis we see that only $C_4$ and $C_5$ are interesting and
  $C_4$ represents late time cosmic acceleration. In 3D analysis,
  the critical points $E_3^\pm$ and $E_7^\pm$ are classically stable but they are
  locally unstable in the phase space (Tables 6 and 8).
 Only critical points
  $E_4$~,~$E_5$~ are locally as well as classically stable and they
  are very interesting from cosmological point of view.
  For the critical points $E_4$~and~$E_5$, for some restriction on the
  independent parameters, they will be stable points and they
  correspond to stable model and yields late time attractors. Here, $E_4$~ represents late time cosmic acceleration. So we see that the present model can realise
  late time cosmic acceleration in all cases.
  It may be noted that under same coupled scalar field represented by action (\ref{3}) in standard cosmology,
  late time acceleration can be realised only for special potential and coupling functions for which phase space
  becomes two dimensional. Moreover the critical points which represent late time cosmic acceleration are not
  classically stable \cite{Mahata:2013oza}. Thus we see that coupled scalar field give rich dynamics in RS2 model
  in contrast to standard cosmology.

In 4D analysis we see that the critical point $C_5$ represents inflation whereas in 3D analysis
(where potential  and coupling function have special forms),  early inflation is not even a critical
point. It means that though RS2 braneworld  favours early inflation
and in this model inflation is possible for a wider class of
potentials than in standard cosmology, it is not generic. Hence, a
unified description of  inflation and phantom (coupled scalar
field) is not possible with those potential functions.  It may be
of considerable interest to  choose $\beta$  (\textit{i.e.}, potential and
coupling functions) such that  inflation is a critical point, so
we leave it for future work.

 {\bf Acknowledgement:}\\
The paper is done during a visit to IUCAA, Pune, India. The first
author is thankful to IUCAA for warm hospitality and facility of
doing research works. The first author would also like to thank
Prof. Varun Sahni and Prof. Subenoy Chakraborty for useful
discussions. Finally, the authors wish to thank the anonymous
referees for helpful suggestions which lead to further improvement
of this work.

\end{document}